\documentclass[prl,twocolumn,showpacs,preprintnumbers,amsmath,amssymb,a4paper,10pt]{revtex4-1}

\usepackage{graphicx,amsmath}

\newcommand{\unit}[1]{\,{\rm #1}}
\newcommand{\sub}[1]{_{\rm #1}}
\newcommand{\subup}[1]{^{\rm #1}}
\newcommand{\e}[1]{{\rm e}^{#1}}

\newcommand{\vctr}[1]{{\bf {#1}}}

\newcommand{\ann}[1]{#1^{\phantom{\dagger}}}
\newcommand{\cre}[1]{#1^{\dagger}}

\begin{document}

\title{Field-dependent superradiant quantum phase transition of molecular magnets in microwave cavities}

\author{Dimitrije Stepanenko$^1$}
\author{Mircea Trif$^2$}
\author{Oleksandr Tsyplyatyev$^{3,4}$}
\author{Daniel Loss$^5$}

\affiliation{$^1$Institute of Physics Belgrade, University of Belgrade, Pregrevica 118, 11080 Belgrade, Serbia}
\affiliation{$^2$Laboratoire de Physique des Solides, CNRS UMR-8502, Universit\'e Paris Sud, 91405 Orsay Cedex, France}
\affiliation{$^3$School of Physics and Astronomy, The University of Birmingham, Birmingham, B15 2TT, UK}
\affiliation{$^4$Institut f\"ur Theoretische Physik, Universit\"at Frankfurt, Max-von-Laue Strasse 1, 60438 Frankfurt, Germany}
\affiliation{$^5$Department of Physics, University of Basel, Klingelbergstrasse 82, CH-4056 Basel, Switzerland}

\date{\today}

\begin{abstract}
We find a superradiant quantum phase transition in the model of triangular molecular magnets coupled to the electric component of a microwave cavity field.  The transition occurs when the coupling strength exceeds a critical value which, in sharp contrast to the standard two-level emitters, can be tuned by an external magnetic field.  In addition to emitted radiation, the molecules develop an in-plane electric dipole moment at the transition. We estimate that the transition can be detected in state of the art microwave strip-line cavities containing $10^{15}$ molecules.
\end{abstract}

%
%
\maketitle

{\it Introduction ---}
The superradiant phase of a collection of emitters coupled to common electromagnetic field mode is characterized by a finite number of photons in the ground state of the combined system.  In the model of two-level emitters coupled to a single cavity mode \cite{D54,JC63,TC68,TC69}, the superradiant phase appears when the emitter-field coupling $g$ exceeds some critical value $g\sub{c}$ \cite{HL73,WH73}.  Theoretical and experimental search for the superradiant phase transition has included atoms and molecules coupled to single- and multimode optical cavities, Josephson junction qubits in microwave resonators, as well as ultracold atoms in optical traps \cite{SS11,BGB+10,BMB+11,DR02,LL09}.

According to the no-go theorem \cite{KAH78, BR79,VVM11}, the ground state of any collection of two-level emitters with dipolar coupling to a mode of electromagnetic field does not contain cavity photons.  This result seems to render the superradiant quantum phase transition impossible, and it was extended to the case of many electromagnetic field modes and many levels in Josephson junctions \cite{BR79,VVM11}.  However, the superradiant phase transition was predicted to occur in the interacting emitters as well as in an ensemble of inhomogeneously coupled emitters and many modes \cite{SS11,ZMD+14}.  It was indeed observed in ultracold gases \cite{BMB+11}.  Here, we consider emission from an ensemble of interacting spins, and we are not aware of any extension of the no-go theorem that applies to our case.
\begin{figure}[t]
\includegraphics[width=7cm]{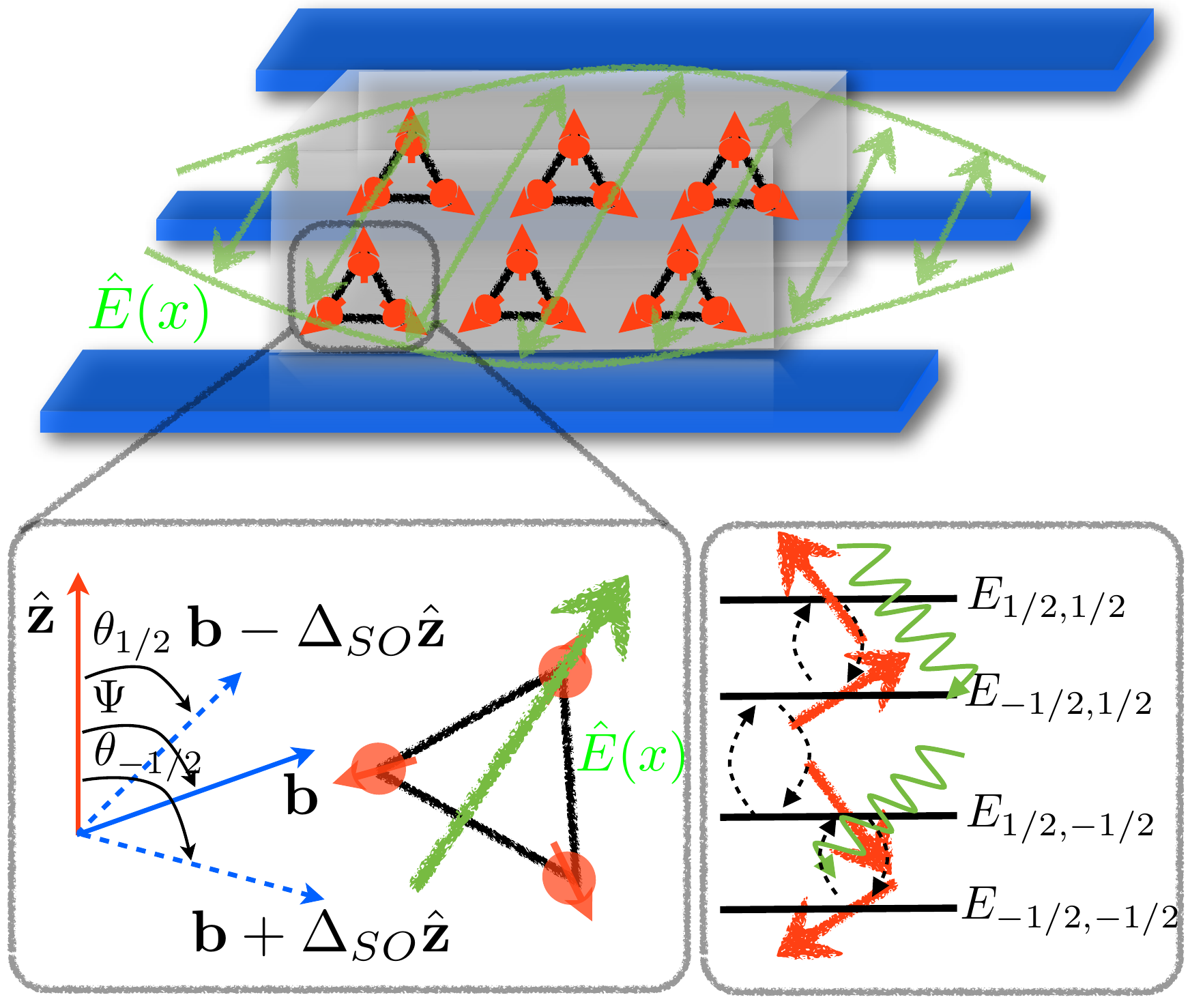}
\caption{\label{fig:setup}Geometry of a crystal of molecular magnets in a microwave cavity and external magnetic field. Electric field of the cavity mode is in the plane of the molecule ($x-y$). External magnetic field $\vctr{B}$ produces the effective fields $\vctr{b}= \mu\sub{B} g\sub{mol} \vctr{B}$, which is tilted by the angle $\psi$ from the normal $\vctr{e}_z$ to the plane of the molecules.  The fields $\vctr{b}_{\pm1/2}$ form angles $\theta_{\pm1/2}$ with the $z$-axis, and define quantization axes of spin (see text).  The angle $\delta=\theta_{-1/2}-\theta_{1/2}$ determines the coupling strength of different transitions.}
\end{figure}

Two-level emitters interacting with the quantized electromagnetic field of resonant cavity are described by the standard Dicke, Jaynes-Cummings, and Tavis-Cummings models of quantum optics \cite{JC63}.  Motivated by the spin-electric coupling of molecular magnets \cite{TTS+08}, we introduce a new model for the emitter in a cavity.  The emitter degree of freedom represents the chirality of ground-state spin texture in a triangular molecular magnet, which interacts with the molecule's total spin.  A crystal with oriented molecular magnets in a strip-line cavity is then described by a generalization of the Dicke model, see Fig.~\ref{fig:setup}. 


We find that the cavity field and molecular magnets can be driven through the transition by modifying the direction or intensity of the external magnetic field.  The critical coupling for the transition is field dependent, due to the interaction between spins within the molecules. Spin interaction makes the ground- and low-energy excited states coherent superpositions of entangled total spin and chirality of the spin texture.  In molecular magnets \cite{GSV06}, the quantum coherence was crucial for explaining the dynamics of magnetization: transitions between the spin states are coherent processes, and show the interference between transition paths \cite{LDG92,LL00b,LML03} and the Berry phase \cite{WS99,GL07,GLM08}.  While our considerations apply to molecular magnets, a range of emitters, like NV centers in diamond, Josephson junctions, N@C$60$ clusters, surface plasmons and excitons in CdSe quantum dots, as well as the rare earth ions may allow for similar field-dependent superradiant quantum phase transition of the interacting spin emitters \cite{TB10,KOB+10,WSB+04,WAB+09,CSH+06,SSW+07,BCG+09}.

{\it Model ---}
At low energy, triangular molecular antiferromagents are characterized by the total spin-$1/2$, $\vctr{S}=\sum_{i=1}^3\vctr{s}_i$, where $i$ counts the spins-$1/2$ on magnetic centers, and pseudospin-$1/2$ chirality $\vctr{C}$, associated with the spin texture, see Fig.~\ref{fig:setup}.  The $z$-component of the chirality is $C_z=\vctr{s}_1\cdot(\vctr{s}_2\times\vctr{s}_3)/(8\sqrt{2})$, and the components $C_x=-(\vctr{s}_1 \cdot \vctr{s}_2 - 2 \vctr{s}_2 \cdot \vctr{s}_3 + \vctr{s}_3 \cdot \vctr{s}_1)/3$ and $C_y=(\vctr{s}_1 \cdot \vctr{s}_2 - \vctr{s}_3 \cdot \vctr{s}_1)/3$ are two-spin operators that flip chirality $C_z$ in analogy with Pauli spin operators \cite{TTS+08}.  The operators $\vctr{S}$ and $\vctr{C}$ are independent and satisfy spin commutation relations: $\left[ S_i, S_j \right] = i \epsilon_{ijk} S_k$,  $\left[ C_i, C_j \right] = i \epsilon_{ijk} C_k$, and  $ \left[ \vctr{S}, \vctr{C} \right] = 0$,  where $i$, $j$, and $k$ count the Cartesian components of spin and chirality \cite{TTS+08,TTS+10}.  

The two degrees of freedom, $\vctr{S}$ and $\vctr{C}$, couple differently to external fields: while the spin couples to the magnetic field via Zeeman term, the chirality couples to the components of external electric field in the plane of the triangular molecule, $\vctr{E}_{\parallel}$\cite{TTS+08}.  The Hamiltonian of the molecular magnet in external electric and magnetic fields is \cite{TTS+08}
\begin{equation}
\label{eq:hmol}
H\sub{mol} = 2 \Delta\sub{SO} C_z S_z + \vctr{b} \cdot \vctr{S} + d_{0} \vctr{E}_{\parallel} \cdot \vctr{C}.
\end{equation}
The Bohr magneton, $\mu\sub{B}$, and the molecular gyromagnetic ratio, $g\sub{mol}$, are absorbed in the effective magnetic field $\vctr{b} = \mu \sub{B} g\sub{mol} \vctr{B}$, and we set $\hbar=1$.  The zero-field splitting, $\Delta\sub{SO}$, caused by the spin-orbit interaction, produces an Ising coupling between $S_{z}$ and $C_{z}$, with the spin $z$ axis normal to the molecule's plane.  In a typical molecular magnet $\Delta\sub{SO}/(g\mu\sub{B})\sim 1\unit{T}$ \cite{GSV06,CMN+06}, setting the control magnetic field strengths to $B\sim 1\unit{T}$, and the resonant frequency of radiation to the microwave region, $\omega \sim 100\unit{GHz}$.  The predicted value of the spin-electric coupling constant is $d_{0}\sim 10^{-4}|eR_0|$ where $e$ is the electron charge, and $R_0$ is the distance between magnetic centers \cite{INC+10,NIC+12}.  The chirality interacts with the in-plane components of the electric field and, through the Ising coupling, with a quantum degree of freedom, $\vctr{S}$ \cite{TTS+08,TTS+10}.

A crystal of $N$ emitters interacting with a mode of the resonant cavity is described by
\begin{equation}
\label{eq:hqjc}
H = H\sub{cav} + \sum_j  H_{0,j} + \sum_j V_{j},
\end{equation}
where $H\sub{cav} = \omega \cre{a}a$ describes the cavity photon, and each $H_{0,j} = 2 \Delta\sub{SO} C_{j,z} S_{j,z} + \vctr{b}\cdot \vctr{S}_j$ describes a molecule interacting with a classical magnetic field $\vctr{B}$.  The interaction terms, $V\sub{j} = d \left( a + \cre{a} \right) C_{j,x}$, are couplings of molecules to the electric component of quantized cavity field.  The operator $a$ ($\cre{a}$) annihilates (creates) a cavity photon.  The coupling constant $d=d_0E_x$ includes both the intrinsic coupling $d_0$ and the electric field amplitude $E_x=\sqrt{\hbar\omega/c_lV}$, where $c_l$ is the resonator capacitance per unit length, and $V$ is the volume of the cavity \cite{BHW+04}.  Assuming the resonant frequencies $\omega$ in the microwave range and state-of-the-art microwave cavities with $E_x\sim 100\unit{V/m}$, we obtain $d\sim10^{-11}\unit{eV}$.  The molecules in a crystal lie in parallel planes, so that their spin quantization axes all point in the same direction, $z$ \cite{BCG+09}.  Any variation of molecular orientations, e.g., due to crystal defects, is equivalent to a change in the effective coupling between the molecular spins and the cavity photons.  The Hamiltonian (\ref{eq:hqjc}) does not contain the Zeeman coupling of spin $\vctr{S}$ to the magnetic component of the cavity field.  This coupling is much weaker than the spin-electric coupling between the electric field and chirality $\vctr{C}$.  Neglecting this term is appropriate for the microwave cavities with molecules placed near the maximum of the electric field amplitude \cite{I09}.

The non-interacting Hamiltonian, $H_0 = H\sub{cav} + \sum_{j}H_{0,j}$, conserves the number of photons $\hat{n} = \cre{a} \ann{a}$, as well as the $z$-components of chiralities, $C_{j,z}$.  Within each simultaneous eigenspace of ${\hat n}$ and $C_{j,z}$ it reduces to a spin Hamiltonian
\begin{equation}
\label{eq:h0ncz}
H_{0,j;n,c}= n \omega + \vctr{b} \cdot \vctr{S}_{j}
+ 2 c \Delta \sub{SO} S_{j,z},
\end{equation}
where $n$ and $c$ are the respective eigenvalues of the operators ${\hat{n}}$ and $C_{j,z}$.  This reduced Hamiltonian is readily diagonalized, and we find the energies,
$E_{n,c,s} = s |\vctr{b}(c)| +  n  \omega$, 
and the eigenstates,
$| n,c,s\rangle  = | n, c\rangle \otimes \left|  \vctr{S}\cdot \vctr{e}_{c} = s \right\rangle$.
The effective magnetic fields are $\vctr{b}(c) =  \vctr{b} + 2 c \Delta\sub{SO} \vctr{e}_z$, with $c=\pm 1/2$.  The eigenstates are $| n , c, s \rangle$, and $s=\pm 1/2$ denotes the molecule's spin projection along $\vctr{e}_c$, the direction of effective field $\vctr{b}(c)$.  Explicitly, the molecule's eigenstates in the $C_{j,z}$, $S_{j,z}$ basis are given by the unitary transformation $|n, c, s \rangle = | n \rangle \otimes U |c, s_z\rangle$, where $U= \sum_{c =\pm 1/2} P_{c} \exp\left( -i \theta_c S_y \right) P_{c}$ maps the state $|c, s_z \rangle$ of the molecule with chirality $c$ and spin projection $s_z$ to the $z$-axis into a state with the same chirality and the spin projection $s= s_z$ along the rotated spin axis (see Fig.~\ref{fig:setup}).  The angles $\theta_{\pm 1/2}$ are
\begin{equation}
\label{eq:thetacz}
\theta_{c} = \arccos \frac{ 2 c \Delta\sub{SO} + b \cos \psi}{\sqrt{ b^2 \sin^2\psi + ( 2 c \Delta\sub{SO} + b \cos\psi)^2}},
\end{equation}
with $\psi$ denoting the polar angle of the field $\vctr{b}$.  The operators $P_{c}=2\, c\, C_z +1/2$ are projectors to the states of a given chirality $c$.

{\it Rotating wave approximation ---}
As opposed to the standard Jaynes-Cummings model in quantum optics \cite{MW95}, the rotating wave approximation (RWA) for a single-molecule magnet in a cavity can not be obtained by simply neglecting the terms proportional to $C_+ \cre{a}$ and $C_{-} a$, since the chirality interacts with the spin, which in addition couples to external fields.
\begin{figure}[t]
\begin{center}
\includegraphics[width=8cm]{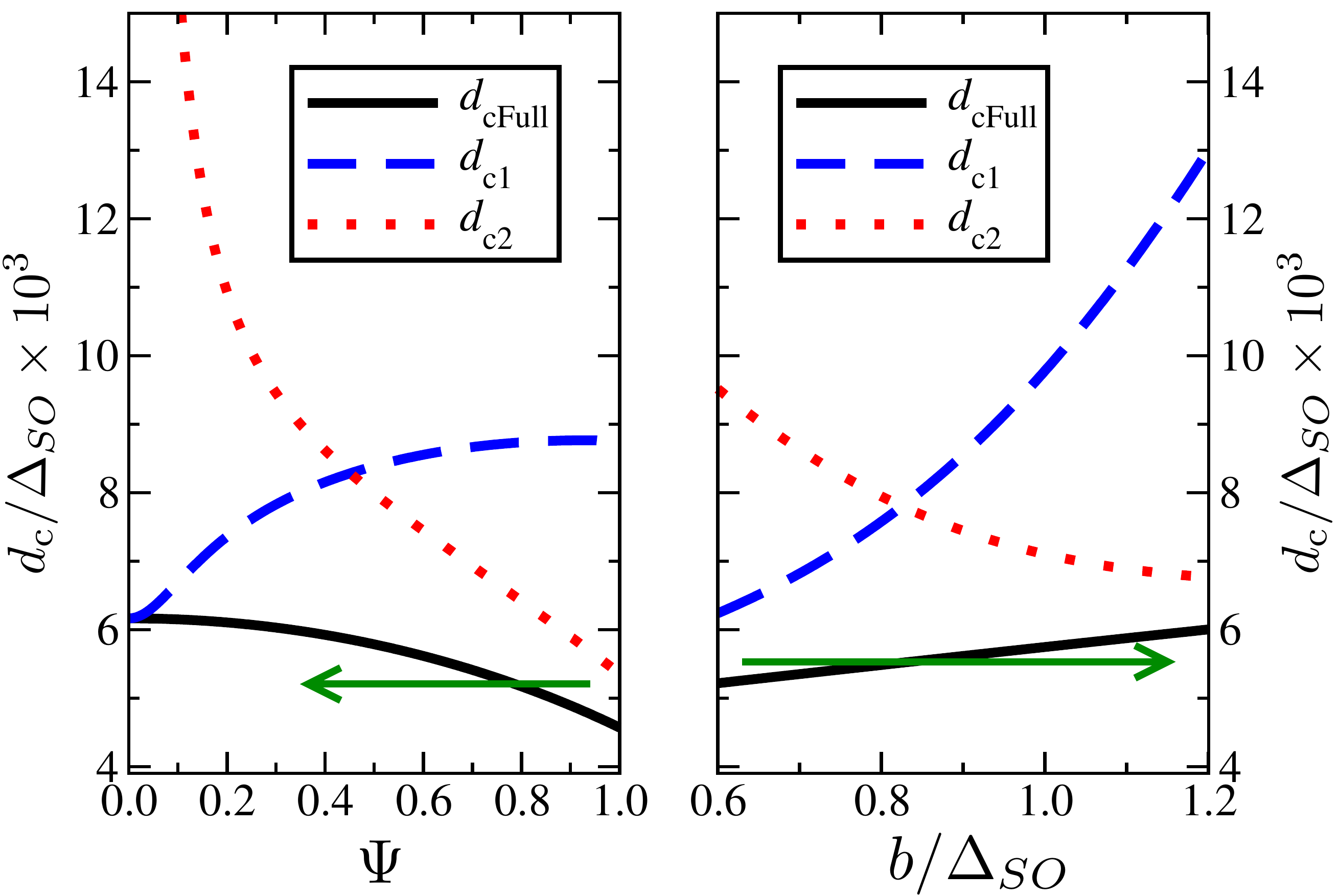}
\caption{
\label{fig:dc}
The critical couplings in the full RWA ($d\sub{cFull}$), in standard  RWAs near $\omega\sub{r}^+$ ($d\sub{c1}$), and near $\omega\sub{r}^-$ ($d\sub{c2}$), as a function of angle with respect to the normal to molecule's plane $\psi$, and the intensity $b$ of the external magnetic field $\vctr{b}$, respectively.  Variations in either $\psi$ or $b$ lead the system through the superradiant quantum phase transition (motion along the arrows switches from $d>d\sub{c}$ to $d<d\sub{c}$).  For this figure, the number of molecules is $N=10^5$, and the cavity frequency is the mean of the two  resonant frequencies $\omega = (\omega\sub{r}^+ + \omega\sub{r}^- )/2$ (see text).  On the first panel $b  = 0.9\Delta\sub{SO}$, and on the second $\psi = 0.6\unit{rad}$.}
\end{center}
\end{figure}

To derive the RWA of Eq.(\ref{eq:hqjc}) we switch to the interaction picture, $V_j(t) =  \e{iH_0 t}V_j \e{-iH_0 t}$, with respect to the terms $H_0 = \sum_j H_{0,j}$ that do not involve the interaction of the molecule with the cavity field.  Using the known eigenvalues and eigenstates of $H_0$, we find
\begin{equation}
\label{eq:vInt5}
\begin{split}
V_j(t) =&\frac{d}{2} \sum_{n,c,s,s'} 
\e{i \left( E_{n,c,s} - E_{n,\bar{c},s'} \right) t} 
M(c,s,s')
\\
& 
|n,c,s\rangle \langle n,\bar{c},s'| 
\left( \e{i\omega t} \cre{a} + \e{-i\omega t} \ann{a} \right),
\end{split}
\end{equation}
where 
$M(c,s,s')=\left\langle \vctr{S}_j \cdot \vctr{e}_{c} = s | \vctr{S}_j \cdot \vctr{e}_{\bar{c}} = s' \right \rangle$ 
is the scalar product of the spins with projections $s$ and $s'$ on the axes $\vctr{e}(c)$ and $\vctr{e}(\bar{c})$, where $\bar{c}=-c$, $\bar{s}=-s$. Explicitly,
$M(c,s,s) = \cos {(\delta / 2)}$,
$M(\pm1/2,s,{\bar{s}})  = \mp i \sin {(\delta / 2)}$,
$\delta=\theta_{-1/2} - \theta_{1/2}$,
and the angles $\theta_{\pm 1/2}$ are given in Eq. (\ref{eq:thetacz}).  

The RWA consists of neglecting the terms in the interaction-picture Hamiltonian (\ref{eq:vInt5}) that oscillate with frequencies close to molecular transitions $\omega_{ij} \sim | E_i - E_j |$, and keeping the terms that oscillate slowly, with frequencies close to the detuning between the transition and the cavity mode.  
In this case the fast-oscillating terms average out to zero, and we can neglect them.  The resonant frequencies in our model are 
$
\omega\sub{r}^{\pm} =
\left( |  \vctr{b}(1/2) | \pm | \vctr{b} (-1/2) | \right)/2. 
$
We have set the direction of $z$ axis so that $ |  \vctr{b}(1/2) | \ge | \vctr{b} (-1/2) |$.

The condition for the validity of the RWA is that the molecule-cavity coupling constant $d$ is much smaller than the resonant frequencies, $d \ll \omega_r^{\pm}$.  In addition, the RWA can reproduce the standard model of a two-level emitter when the cavity frequency is tuned close to one of the transitions and far from the other, e.g., $|\omega - \omega^{+}|\gg |\omega - \omega^{-}|$.  This tuning is possible only when
\begin{equation}
\label{eq:singleresonance}
|\omega\sub{r}^{+} - \omega\sub{r}^{-} |  \gg d.
\end{equation}
The condition (\ref{eq:singleresonance}) can not be satisfied when $\vctr{b}\approx \Delta\sub{SO}\vctr{e}_z$, i.e., when the magnetic field axis is near the normal to the molecule, and the magnetic field intensity is comparable to spin-orbit splitting $\Delta\sub{SO}$ (usually around $1\unit{T}$ \cite{CMN+06}).  We will focus on the case when both resonances have to be taken into account, either due to the deliberate tuning of the cavity frequency, or due to violation of Eq. (\ref{eq:singleresonance}).  In this case, the amplitudes of the resonant transitions vary strongly with the magnetic field, and we will see that this leads to new effects.  When Eq. (\ref{eq:singleresonance}) is satisfied, the cavity can be tuned so that the RWA leads to the Tavis-Cummings model \cite{TC68,TC69}, and consequently to the familiar superradiant phase transition and a single transition resonant with the cavity, see Fig.~{\ref{fig:dc}}.

After the removal of the counter-rotating terms and switching back to the Schr\"odinger picture, the molecule-cavity interaction is
\begin{equation}
\label{eq:vrwa}
\begin{split}
V\sub{RWA} =& d \sum_j
\left( a + \cre{a} \right) \left( \frac{\cos\delta}{2} C_{j,x} - \sin\delta\, S_{j,y}C_{j,y} \right)
\\
+ i 
&
\left( a - \cre{a} \right) \left( \sin\theta_{-\frac{1}{2}} S_{j,x} + \cos\theta_{-\frac{1}{2}} S_{j,z} \right) C_{j,y}
.
\end{split}
\end{equation}
The final Hamiltonian in RWA is
$
H\sub{RWA}=H_0 + V\sub{RWA}
$,
and it is analogous to the Tavis-Cummings model of two-level atoms in a resonant cavity.  Similarly to the conservation of the number of excitations in the Tavis-Cummings model, $H\sub{RWA}$ conserves the quantity 
$N\sub{exc} = \hat{n} + \sum_j \left( 1 + \tilde{S}_{j,z} + 2 C_{j,z} \tilde{S}_{j,z} \right)$,
where $\tilde{\vctr{S}}_j = U \vctr{S}_j U^{\dagger}$, with $U$ defined above Eq. (\ref{eq:thetacz}).  The number of excitations, $N\sub{exc}$, is conserved if we count molecules in the state $|c,s\rangle=|1/2,-1/2\rangle$ as zero excitations, molecules in the states $|-1/2,\pm 1/2\rangle$ as one excitation, molecules in the state $|1/2,1/2\rangle$ as two excitations, and each cavity photon as one excitation.  We choose an additive constant so that $N\sub{exc}=0$ corresponds to all the molecules in the state $|1/2,-1/2\rangle$ and no photons in the cavity.

{\it Superradiant quantum phase transition ---}
We study the superradiant phase transition in the rotating wave and mean-field approximations.  This amounts to substituting photon annihilation(creation) operator $a$($\cre{a}$) by their expectation value $\langle a \rangle$($\langle a \rangle ^*$), thus neglecting any quantum fluctuations.  This approximation is valid for large photon numbers.  We find the minimum of the ground state energy of $H\sub{RWA} = \omega|\langle a \rangle |^2 + \sum_{j} H_{0,j} + V\sub{RWA}(\langle a \rangle)$ as a function of $\langle a \rangle$.  The critical coupling is $d\sub{c}$, the smallest value of $d$ for which the minimum lies at $|\langle a \rangle|=\langle a \rangle\sub{MF}>0$.  The mean-field energy is independent of the phase of $\langle a \rangle$, which we set to be real in further discussion. Without RWA, $\langle a \rangle\sub{MF}$ is real \cite{BC14}.  The value of $\langle a \rangle\sub{MF} $ is zero for $d<d\sub{c}$, and increases, $ \langle a \rangle \sub{MF} \propto \sqrt{d-d\sub{c}}$ for $d>d\sub{c}$.  In mean-field $\langle a \rangle \sub{MF} \propto \sqrt{N}$.
\begin{figure}[t]
\begin{center}
\includegraphics[width=8cm]{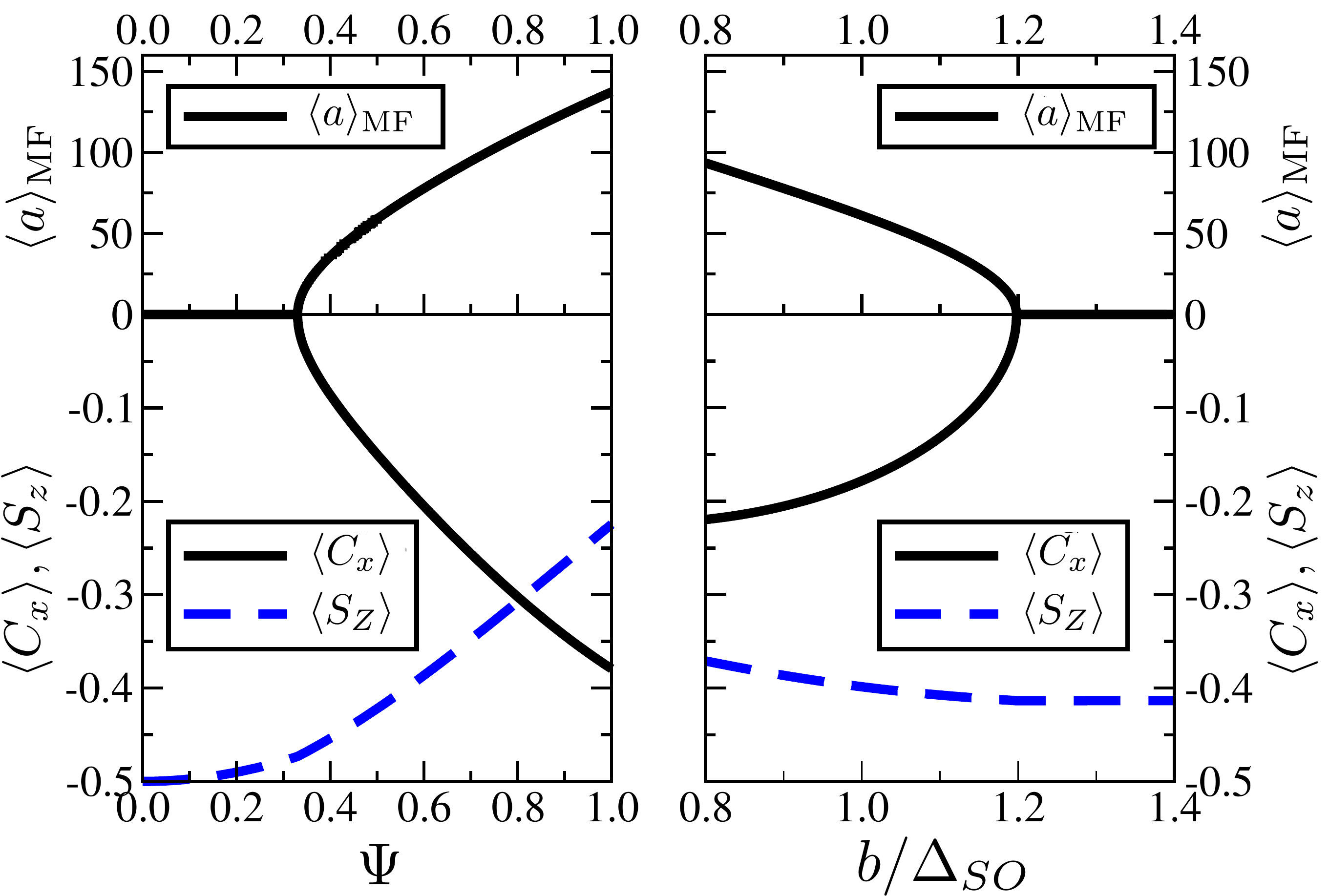}
\caption{
\label{fig:responseB}
Response of molecules and cavity field to the changes in direction $\psi$ (first panel) and intensity $b$ (second panel) of the external magnetic field $\vctr{b}$.  At the superradiant transition, the mean-field value of the photon annihilation operator $\langle a \rangle \sub{MF}$ becomes nonzero (upper panels).  At the same value of $\vctr{b}$, an in-plane electric polarization $\propto \langle C_x \rangle$ appears, signaling the superradiant phase.  The magnetization normal to the molecule's plane $\propto \langle S_z\rangle$ shows a more rapid change with $\vctr{b}$ then in the normal state.  System parameters are the same as in Fig.~\ref{fig:dc}, and $d=6\cdot 10^{-3}\Delta\sub{SO}$. }
\end{center}
\end{figure}

The critical coupling is determined as the value of $d$ for which $\partial ^2_{\langle a \rangle} E\sub{MF}(\langle a \rangle) |_{\langle a \rangle =0} < 0$.  Since $E\sub{MF}(0)$ is finite, $\lim_{\langle a \rangle \to \infty} E\sub{MF}= \infty$, and $\partial _{\langle a \rangle} E\sub{MF}|_{\langle a \rangle =0}=0$, this condition guarantees the existence of a minimum for the mean-field energy that is lower than $E\sub{MF}(0)$ at some finite value of $\langle a \rangle \sub{MF}$.  Taking both resonances into account, Eq. (\ref{eq:vrwa}), we find
\begin{equation}
\label{eq:dcfull}
d\sub{cFull}  = \sqrt{\frac{8\omega\Delta\sub{SO}b}{N\left[ b(\frac{1}{2}) +b(-\frac{1}{2}) \cos\delta \right]}}.
\end{equation}
This $\vctr{b}$-dependent $d\sub{cFull}$ is one of our main results, Fig.~{\ref{fig:dc}}.  The dependence is due to both the modification of the energy levels of $H_0$, and to modification of the coupling constants for transitions through spin-overlap terms in Eq. (\ref{eq:vrwa}).  The result, Eq. (\ref{eq:dcfull}) clearly can not be explained by the usual RWA at either of the resonant frequencies, as illustrated in Fig.~{\ref{fig:dc}}. 

The dependence of $d\sub{c}$ on $\vctr{b}$ allows for a controllable superradiant phase transition.  Changes in $d\sub{c}$, given by Eq. (\ref{eq:dcfull}), can lead the system into or out of the superradiant phase, see Fig.~\ref{fig:dc}.  The measurement of the escaping radiation as done, for example, by using input-output theory \cite{CDG+10}, would then serve as a signature of superradiant state \cite{CBC05,BMB+11,AB12,KEB13}.  In addition to the nonzero photon occupation of the cavity mode, see Fig.~\ref{fig:responseB}, the transition is characterized by a change in the expectation value of the chirality.  For $d<d\sub{c}$, the molecules are in the state with $C_{j,z}=-1/2$, with zero expectation values of $C_{j, x(y)}$.  After the transition, for $d>d\sub{c}$, the in-plane components of chirality have nonzero expectation value, i.e., $\langle C_{j,x}\rangle \neq 0$ in our model.  The fact that only the $x$-components gets a finite expectation value comes from our phase convention for $\langle a \rangle$ \cite{BC14}.  The molecules develop electric dipole moments for $d>d\sub{c}$, and the transition can be detected by the electric response, as well as by the emitted radiation, lower panels of Fig.~{\ref{fig:responseB}}.  An alternative way to control the transition is to deform the cavity, and therefore change the amplitude of electric field $E_0$ and $d\propto E_0$.

Experiments which would allow one to detect the controllable superradiant phase transition and the spin-electric coupling are within reach of current state-of-the-art.  In the strip-line microwave cavities, the electric field amplitude can be of the order of $E_0\sim100\unit{V/m}$ \cite{TGL08}.  The predicted molecular spin-electric coupling constant is estimated at $d\sub{mol}\sim 10^{-4}|eR_0|$ \cite{INC+10,NIC+12}, where $R_0$ is the distance between magnetic centers, of the order $5\cdot 10^{-10}\unit{m}$ \cite{BCG+09}, and $e$ is the electron charge.  The transition occurs when $d=d\sub{mol}E_0 > d\sub{c}$.  According to Eq. (\ref{eq:dcfull}), this is satisfied in crystals containing $N\sim 10^{15}$ molecules with a typical $\Delta\sub{SO}\sim 1K$.  

The disorder in the molecule's energies due to imperfections of the crystal may bring some of the molecules out of resonance and reduce the effective $N$ below the total number of molecules.  However, the superradiant effect also suppresses such inhomogeneous broadening \cite{TL09,TL10,STF12}.  When the collective coupling of many emitters exceeds the bandwidth of their ensemble, the broadening vanishes altogether so that even far off-resonant molecules interact strongly with the field mode.  This allows one to increase the number of active emitters in the cavity in realistic devices.

{\it Conclusions---}
We have introduced a model of a crystal of single-molecule triangular antiferromagnets interacting with an external classical homogeneous magnetic field and the electric component of a quantized cavity field.  The model shows a superradiant quantum phase transition with the critical coupling tunable by applied magnetic field.   The strong coupling regime is characterized by nonzero mean photon number and electric dipole moment in the triangle plane.  With state-of-the-art cavities and current estimates of spin-electric coupling strength, the tunable transition is achievable for $10^{15}$ molecules in the cavity.  While our models describes single-molecule magnets, it may also be useful in the study of other emitters described by entangled discrete degrees of freedom.

{\it Acknowledgments}: We acknowledge discussions with Filippo Troiani.  This work is funded from Serbian MPNTR grant OI171032, Swiss NF through NCCR QSIT and SCOPES IZ73Z0152500, public grant from the Laboratoire d'Excellence Physics Atom Light Matter (LabEx PALM, reference: ANR-10- LABX-0039), and EPSRC  Grant No. EP/J016888/1.

\bibliography{all}
\bibliographystyle{apsrev}

\end{document}